\documentclass[11pt,twoside]{article}
\usepackage{asp2004}
\usepackage{psfig}
\usepackage{epsf}
\usepackage{graphics}
\usepackage{lscape}
\newcommand{\ea}{et al.}

\newcommand{\kpc}{\>{\rm kpc}}

\newcommand{\msun}{\>{\rm M_{\odot}}}
\newcommand{\lsun}{\>{\rm L_{\odot}}}

\newcommand{\bdm}{\begin{displaymath}}
\newcommand{\edm}{\end{displaymath}}
\newcommand{\beq}{\begin{equation}}
\newcommand{\eeq}{\end{equation}}
\newcommand{\bit}{\begin{itemize}}
\newcommand{\eit}{\end{itemize}}
\newcommand{\ben}{\begin{enumerate}}
\newcommand{\een}{\end{enumerate}}
\newcommand{\bfi}{\begin{figure}[htb]}
\newcommand{\bpfi}{\begin{figure}[p]}

\def\edcomment#1{\iffalse\marginpar{\raggedright\sl#1\/}\else\relax\fi}
\reversemarginpar

\begin{document}
\title{The Nuclei of Late-type Spiral Galaxies}
\author{T. B\"oker$^1$, C.-J. Walcher$^2$, H.-W. Rix$^2$, N. H\"aring$^2$, 
	E. Schinnerer$^2$, M. Sarzi$^3$, R. P. van der Marel$^4$, L. C. Ho$^5$,
	J. C. Shields$^6$, U. Lisenfeld$^7$, S. Laine$^8$}
\affil{$^1$European Space Agency (ESTEC), 2200 AG Noordwijk, Netherlands \\
       $^2$MPI f\"ur Astronomie, K\"onigsstuhl 17, D-69117 Heidelberg, Germany \\
       $^3$University of Oxford, Keble Rd., Oxford OX1 3RH, U.K. \\
       $^4$STScI, 3700 San Martin Drive, Baltimore, MD 21218, U.S.A. \\
       $^5$Carnegie Observatories, 813 Santa Barbara Street, Pasadena, CA 91101-1292,
            U.S.A. \\
       $^6$Ohio University, Dept. of Physics and Astronomy, Athens, OH 45701-2979,
       U.S.A. \\
       $^7$IAA, Camino Bajo de Huetor 24, 18080 Granada, Spain \\
       $^8$Spitzer Science Center, CalTech, Pasadena, CA 91125, U.S.A.}
%
%
%
%
%
%
%
%
\begin{abstract}
We summarize some recent results from our observational campaign to
study the central regions of spiral galaxies of late Hubble type 
(Scd - Sm). These disk-dominated, bulgeless galaxies have an
apparently uneventful merger history. The evolution of their nuclei 
thus holds important constraints on the mechanisms that are responsible 
for bulge formation and nuclear activity in spiral galaxies. 
We discuss the structural properties, stellar populations, and dynamical
masses of the compact, luminous star cluster that is found in the
nuclei of most late-type spiral galaxies. Although preliminary, our
results strongly indicate that many galaxies of our sample experience
repeated periods of nuclear star formation. While the exact mechanism 
that leads to the required high gas densities in the galaxy nucleus 
remains unclear, results from our recent CO survey of late-type spirals
demonstrate that in most cases, the central kpc contains enough molecular
gas to support repetitive nuclear starbursts.
\end{abstract}
\section{Introduction: the Central Regions of Late-type Spirals}
In most formation scenarios for spiral galaxies, the central bulge is 
the ``trashbin of violent relaxation'' where a dynamically hot stellar 
component has formed either through external potential perturbations 
such as early mergers of proto-galaxies (e.g. Carlberg 1992), or perhaps 
via internal effects such as violent bar instabilities (e.g. Norman, Sellwood,
\& Hasan 1996). 
The latest-type spirals, then, must have lived very sheltered and uneventful
lives, since their central ``trashbin'' is virtually empty, i.e. they
are devoid of prominent starburst events, have no discernible stellar
bulges, and rarely show signs of nuclear activity. These galaxies often 
have gently rising rotation curves (e.g. Matthews \& Gallagher 1997) that 
indicate a nearly homogeneous mass distribution on scales $\sim 1\kpc$. 
On these scales, gravity therefore hardly provides
a vector pointing at the center, and it is not obvious that the nucleus
of these galaxies is well-defined and a unique environment.

\begin{figure}[!ht] 

\vspace*{4cm}
\centerline{see attached file fig1a.gif} 
\plotone{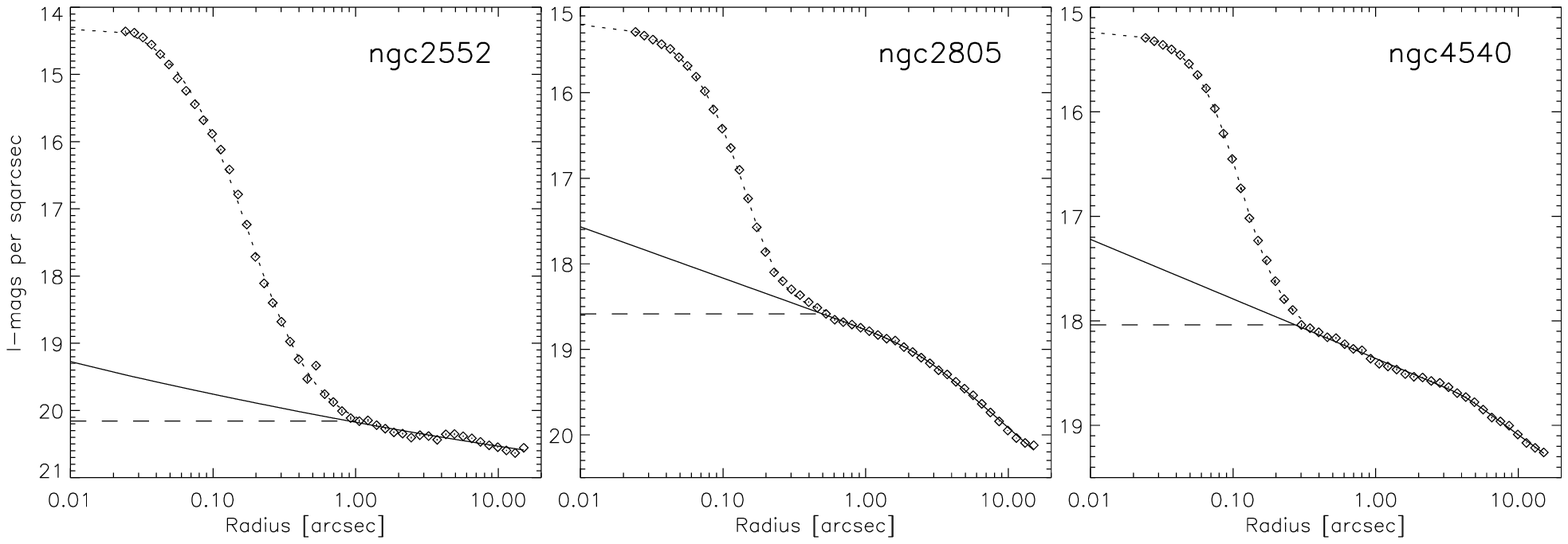}
\caption{Top: HST/WFPC2 F814W (I-band) images of three representative 
nuclear star clusters. Shown is the PC chip with a field of view of 
$\approx 35\arcsec \times 35\arcsec$. The bar in the top left of each panel 
denotes a spatial scale of 1 kpc, the north-east orientation is 
indicated by the compass arrow. Bottom: I-band surface brightness 
profiles, measured from elliptical isophote fits to the images above.
Note the clear transition between the underlying disk and the NC at radii
around $0.2\arcsec$. Also shown are analytical fits to the disk and cluster
profiles that yield the cluster photometry (for details, see B\"oker et 
al. 2002.)}
\label{fig:images}
\end{figure}
Surprisingly, the photocenter of many late-type spiral galaxies 
is nevertheless occupied by a compact, luminous stellar cluster
(Phillips et al. 1996; Carollo et al. 1998; Matthews et al. 1999;
B\"oker et al. 2002, hereafter Paper~I). Figure~\ref{fig:images} shows HST/WFPC2
I-band images of three representative examples of such nuclear star 
clusters (NCs), together with their surface brightness profiles as 
measured from elliptical isophote fits to the WFPC2 data. The typical 
luminosities of NCs are in the range $10^6$ - $10^7\,\lsun$ (Paper~I). 
Nuclear clusters are therefore much brighter than average stellar 
clusters in the disks of nearby spiral galaxies (e.g. Larsen 2002), 
and comparable to young ``super star clusters'' in luminous merger 
pairs (Whitmore et al. 1999) or circumnuclear 
starforming rings in spiral galaxies (e.g. Maoz et al. 2001). 

Knowledge of the stellar populations and masses of NCs is essential
in order to constrain the mechanism(s) that lead to their formation. 
The age(s) of the stellar population(s) can be determined from
spectral synthesis methods applied to medium-resolution spectra 
of NCs. Measuring the stellar masses of NCs requires both accurate 
knowledge of the velocity dispersion (from high-resolution 
spectra) and the cluster light distribution (from HST imaging). 
This kind of analysis has been successfully applied to NCs 
(B\"oker et al. 1999) as well as young clusters in the disks of nearby 
galaxies (Smith \& Gallagher 2001; Mengel et al. 2002, McCrady et al. 2003).
In this paper, we summarize our ongoing study of the properties 
and formation mechanism(s) of NCs in late-type spirals. 
\section{The Sizes of Nuclear Clusters }
Star clusters in all but the closest galaxies are marginally resolved
sources, even with the excellent resolution of HST images. 
In this situation, conventional deconvolution methods 
are not reliable. 

Recently, a number of authors (e.g. Kundu \& Whitmore 1998, Larsen 1999,
Carlson \& Holtzman 2001)
have independently developed techniques to determine more reliably the
structure of stellar clusters in external galaxies. The various
methods are conceptually similar in that they assume a parametric
model for the intrinsic light distribution. 
A particular model is considered to be a good description of reality 
if - after convolution with the instrumental PSF - it matches the 
observations. The ``best'' model is the one that minimizes the 
$\chi^2$-difference to the data. 

\begin{figure}[!ht]  
\plottwo{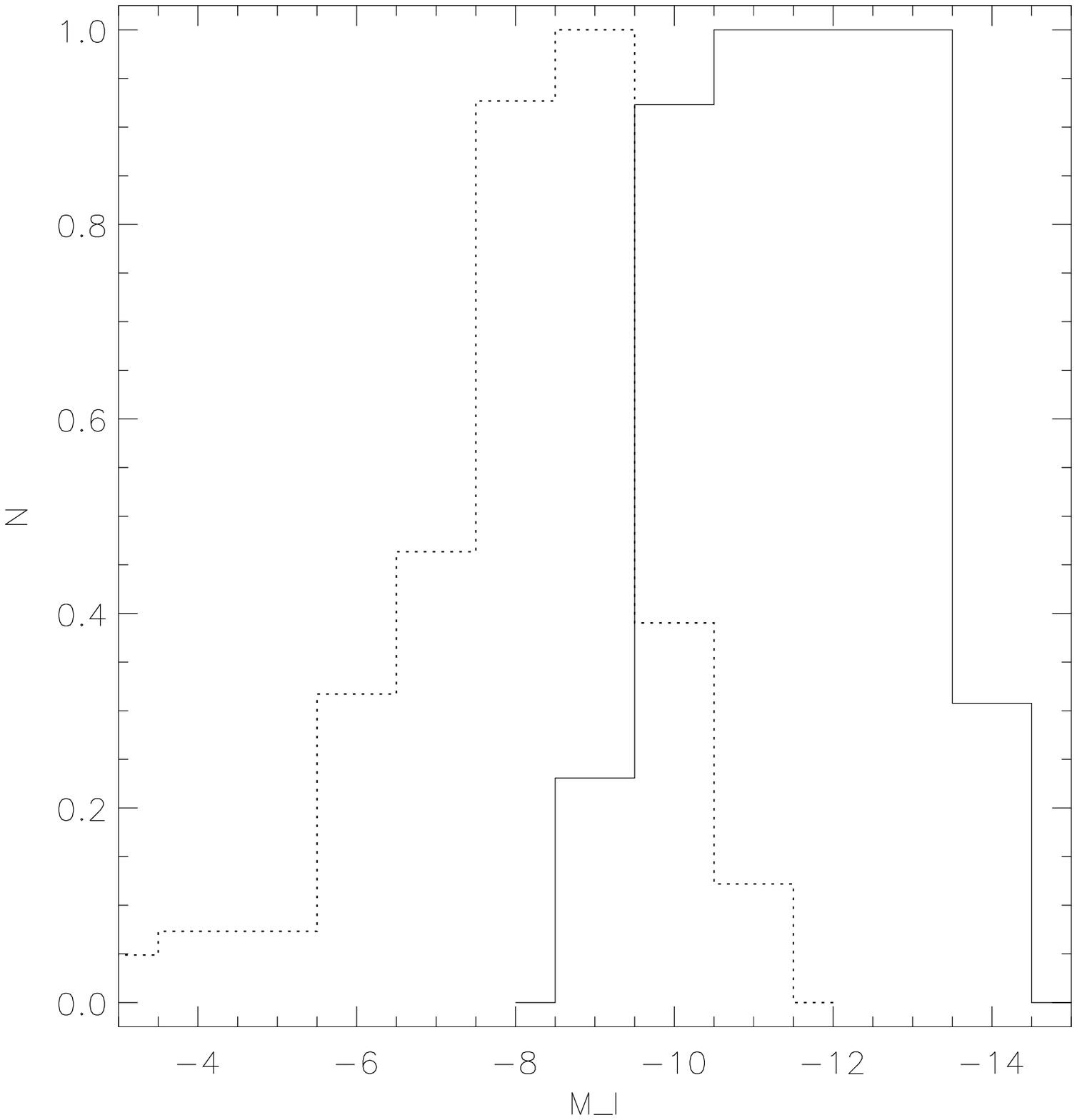}{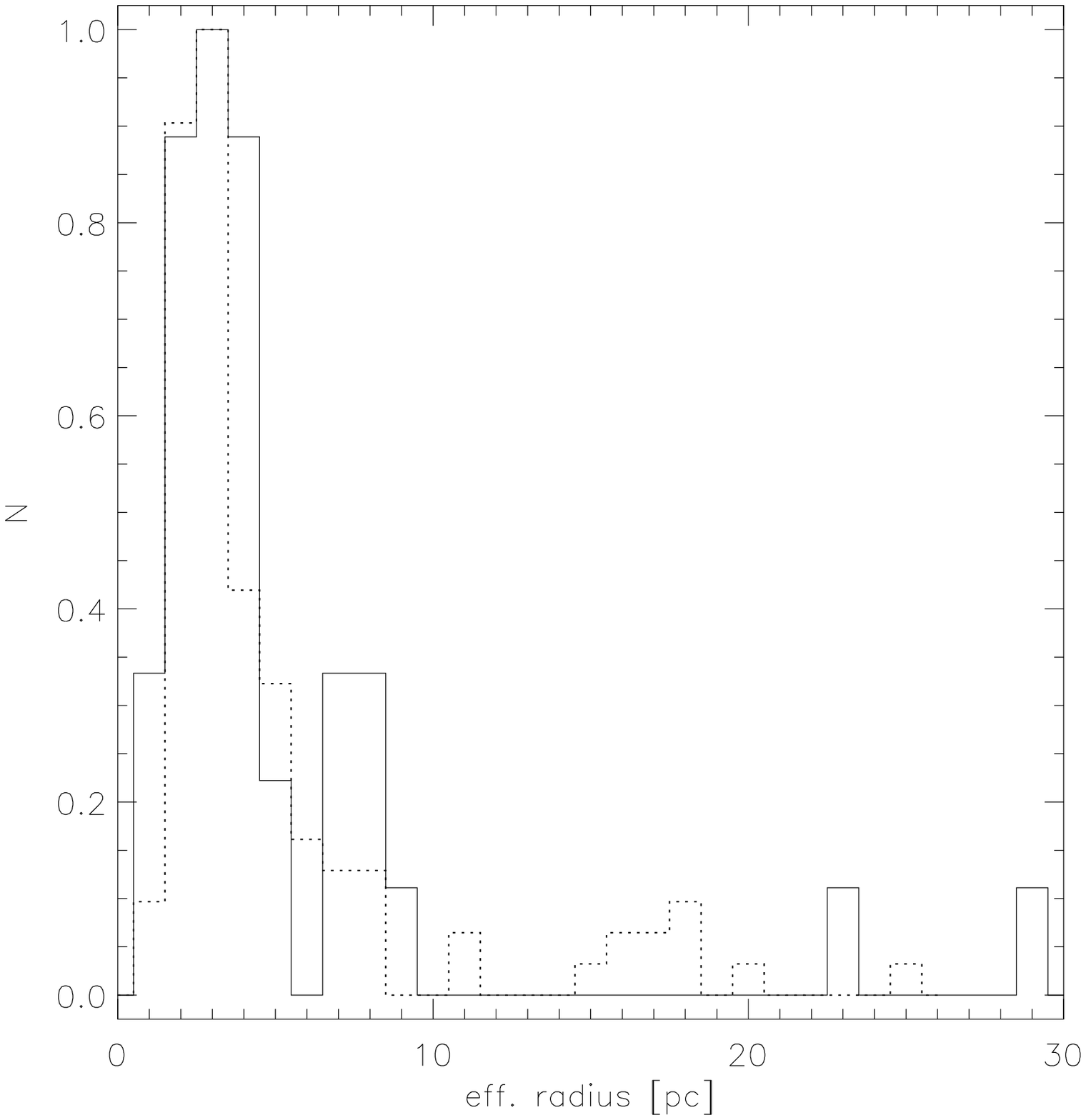}
\caption{Luminosity (left) and size (right) distribution of nuclear
star clusters, as compared to Milky Way globular clusters (dotted lines).
Both families of clusters have similar sizes, but nuclear clusters are
on average 4 magnitudes brighter (B\"oker \ea\ 2004).}
\label{fig:hists}
\end{figure}

For our analysis, which is described in detail in B\"oker et al (2004,
hereafter paper~II), we used the software package ISHAPE (Larsen 1999)
in combination with synthetic PSFs created with the TinyTim 
software (Krist \& Hook 2001). Unless the signal-to-noise ratio of the 
data is exceptionally high, such an analysis normally identifies a number
of analytical models that are consistent with the data in a $\chi^2$ sense,
and the exact shape of the cluster cannot be constrained further.
However, the cluster size, as measured by the effective radius 
$r_{\rm e}$ (i.e. the radius which includes half of the cluster light), 
is a rather robust quantity, i.e. different acceptable models yield 
very similar values for $r_{\rm e}$.

The results are summarized in Figure~\ref{fig:hists} which shows the luminosity and size
distributions of the nuclear cluster sample, compared to the Milky Way
globular clusters (MWGCs) from the online version of the Harris (1996) 
catalog. While NCs are on average four magnitudes brighter than MWGCs, they
have very similar sizes, with typical values of $r_{\rm e}$ in the range
2 - 5 pc. This is an interesting result because it provides support
for theories which invoke galaxy cannibalism in the early universe
to explain the properties of some unusual GC such as Omega Cen.
Provided that the progenitor galaxies had NCs
similar to ones observed in our sample, these very dense clusters would
most likely have survived the merger and - being stripped of their parent
galaxies - would passively age in the halo of the remnant. For this to be
a viable scenario, one would have to postulate that the 4 magnitudes of
luminosity difference mostly reflect a difference in the stellar
population ages between present-day NCs and MWGCs. In principle, this
is entirely possible: population synthesis models such as PEGASE
(Fioc \& Rocca-Volmerange 1997) or Starburst99 (Leitherer et al. 1999)
predict a fading of 3-4 magnitudes 
in I-band for single age populations between $10^7$ and $10^{10}$ years.

In order to test whether NCs in present-day galaxies are indeed dominated
by young stellar populations, we have begun a spectroscopic follow-up
to determine their ages and dynamical masses.

\section{Stellar Populations of Nuclear Clusters}

\begin{figure}[!ht]  
\plotfiddle{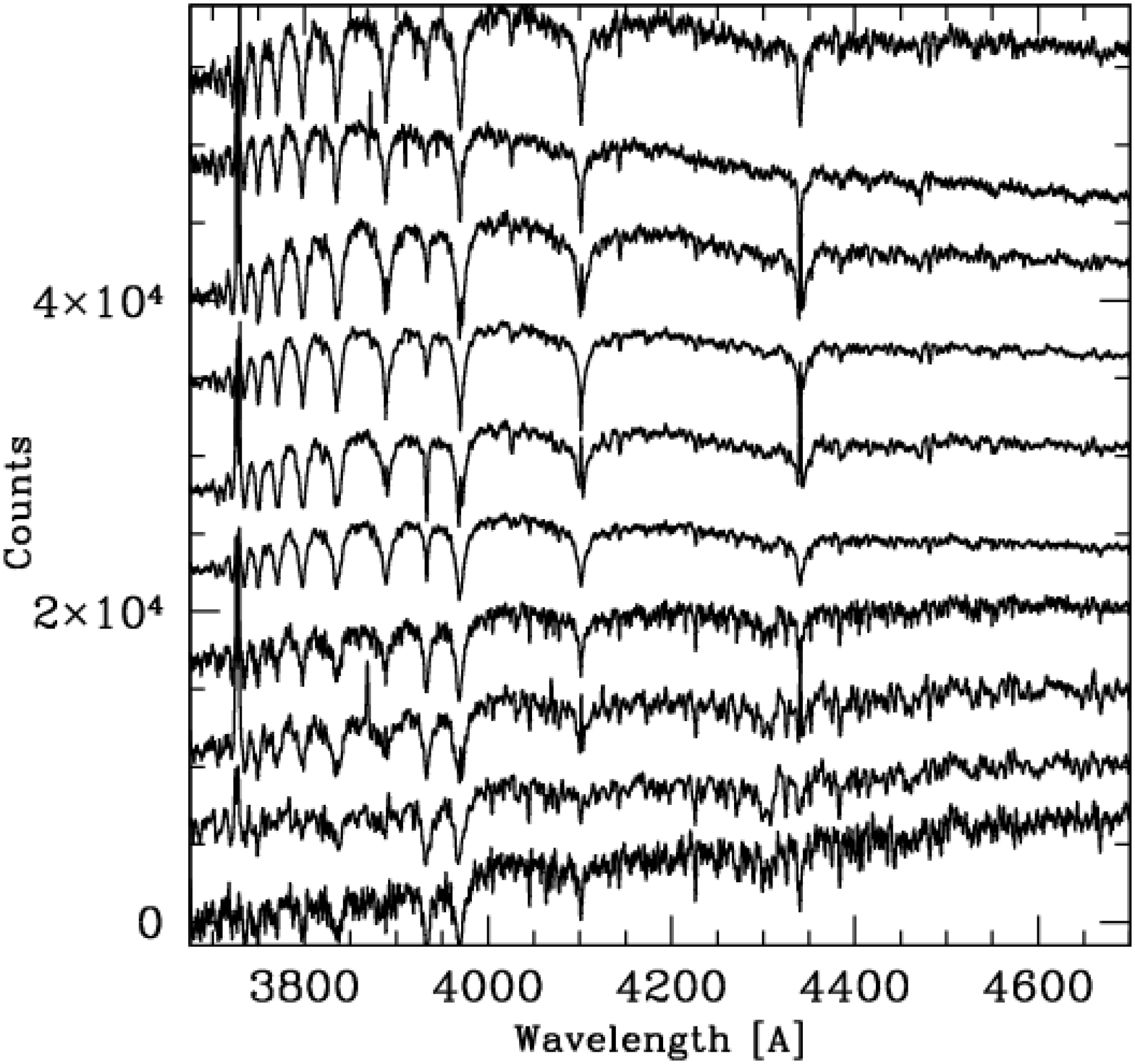}{8cm}{0}{45}{30}{-180}{0}
%
\caption{VLT/UVES spectra of the nuclear star cluster in (from top to bottom)
NGC\,1493, NGC\,2139, NGC\,7424, NGC\,1385, NGC\,7418, NGC\,7793, NGC\,300, 
NGC\,1042, NGC\,3423, and NGC\,428. Note the prominent Balmer absorption 
lines in many clusters which indicate a rather young age ($\leq 1\,$Gyr) 
of the dominant stellar population (Walcher \ea\ 2004, in prep.).
}
\label{fig:spectra}
\end{figure}

Figure~\ref{fig:spectra} shows the "blue arm" optical spectra of 10 nuclear 
clusters taken with VLT/UVES after subtraction of the emission from the 
galaxy disk underlying the cluster. The spectra have a resolution of 
$\rm R\approx 30000$ but have been smoothed for plotting.
The 10 clusters were sorted "by eye" in an age sequence, starting at 
the top with the youngest objects that have the strongest Balmer series 
absorption lines, to the oldest objects with the strongest Ca H+K lines
at the bottom.

It is clear even from this crude first order age-dating attempt that
many clusters show strong Balmer absorption lines which occur only in 
stars of spectral type B or A. These stars must have formed rather 
recently, because they only live for less than 1 Gyr.
Because it is unlikely that a large fraction of nuclear clusters experience
a star formation event at the same time, one can already deduce from
Figure~\ref{fig:spectra} that nuclear starbursts are likely a repetitive event. 

\begin{figure}[!ht]  
\plotfiddle{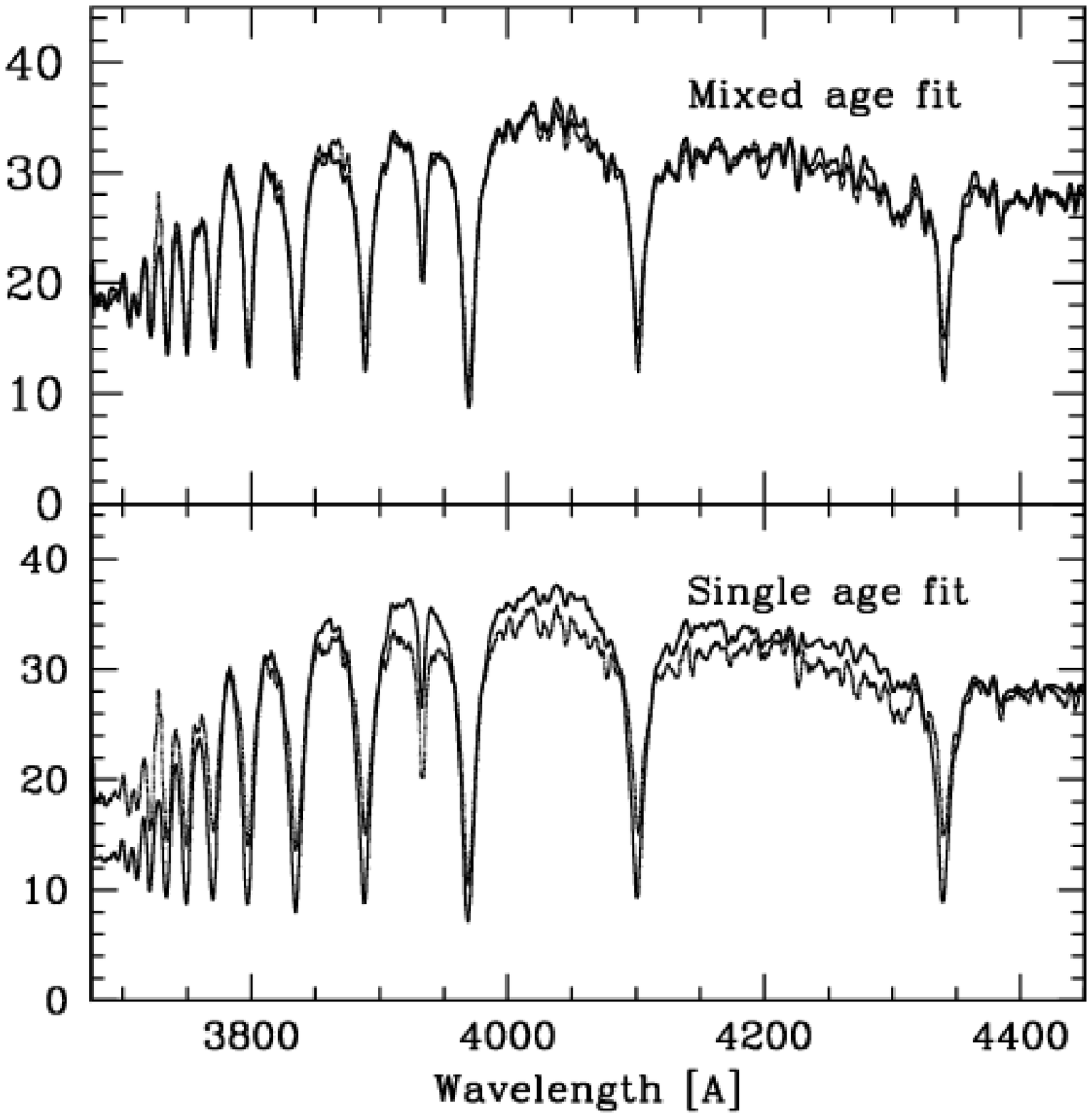}{9cm}{0}{70}{50}{-200}{-14}
\caption{Preliminary population synthesis fitting of the NC in NGC\,7793
(Walcher et al 2004, in prep.).
Bottom: best single age (100 Myrs) fit. Top: best fit using a mix of 
single-age templates ranging from 1 Myr to 10 Gyr.
}
\label{fig:popsynth}
\end{figure}

In fact, this notion is supported by a detailed analysis of the spectra 
in Figure~\ref{fig:spectra} with
population synthesis tools. While this work is currently ongoing, and 
results will be published in a forthcoming paper (Walcher et al. 2004), we
show in Figure~\ref{fig:popsynth} the representative example of NGC\,7793. 
The spectrum can 
not be well described with a single-age stellar population: additional older
populations are required to explain the continuum shape and the presence of 
deep Ca H+K features in an otherwise "young" spectrum.

\begin{figure}[!ht]  
\plotfiddle{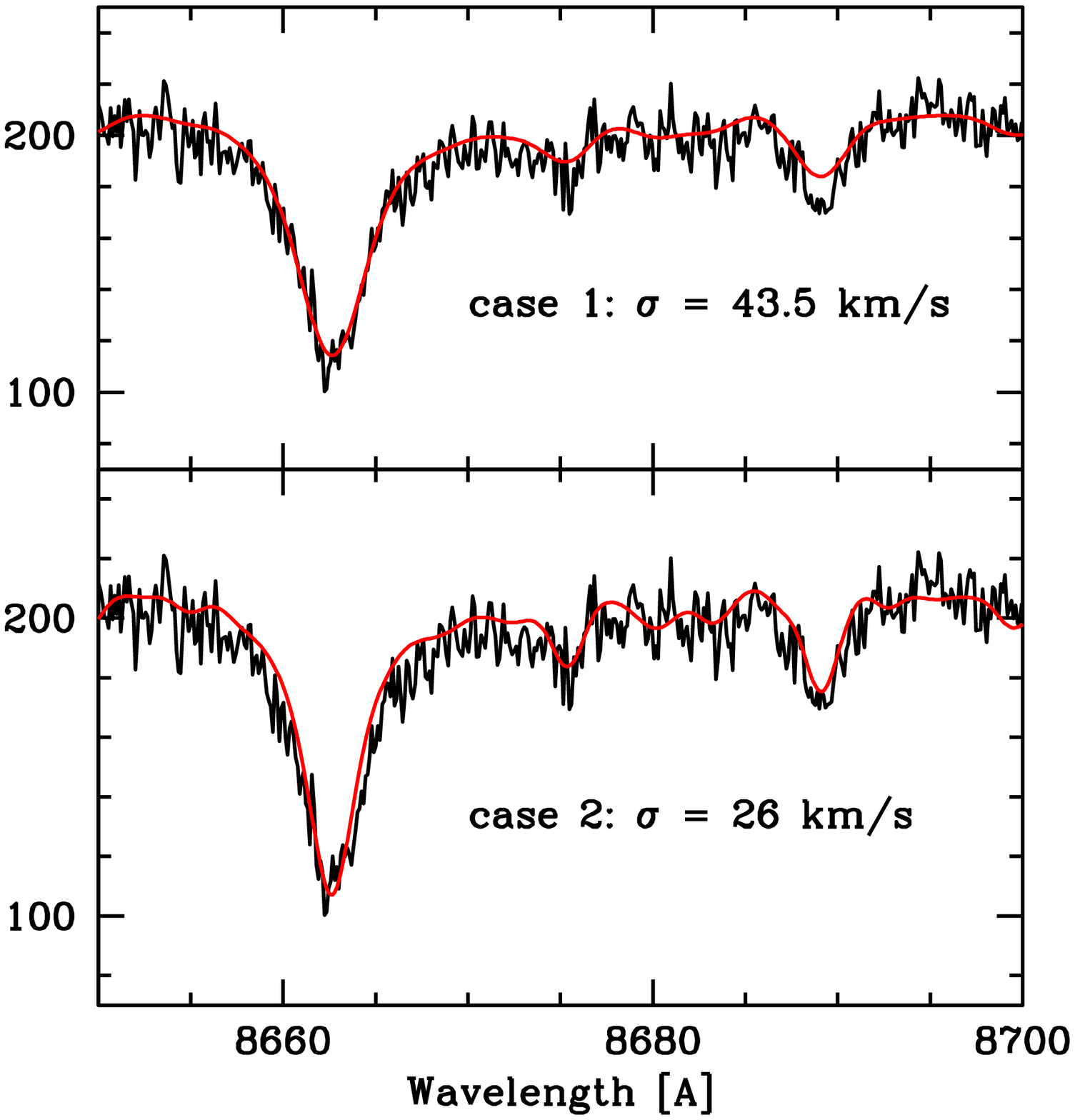}{9cm}{0}{50}{50}{-160}{-70}
\caption{Preliminary results of the dynamical modeling of the NC in
NGC\,1042 (Walcher et al 2004, in prep.). Top: best fit over the entire 
spectral region shown in the plot. 
The fit is dominated by the Ca line at 8662\AA which - because of the 
presence of supergiants - results in too high a value for $\sigma$. 
Bottom: fit restricted to the metal lines, resulting in a lower value
for $\sigma$.
}
\label{fig:cat}
\end{figure}
\section{Dynamical Masses of Nuclear Clusters}
The ``weighing'' of stellar clusters from their
stellar dynamics requires two observational ingredients: 
the stellar velocity dispersion $\sigma$ and an accurate
knowledge of the cluster light distribution (after correction
for any instrumental broadening). For the latter, we
use the results of the ISHAPE fitting described in \S~2.

The stellar velocity dispersion can, in general, be measured 
from the comparison of absorption line profiles in the 
integrated cluster light with those of suitable template 
stars. To first order, any line broadening is assumed to be
due to the Doppler shifts of stars moving in the gravitational
potential of the cluster. Finding the potential that correctly 
predicts $\sigma$ (via Jeans' equation) {\bf and} the cluster
light profile yields the mass-to-light ratio, and hence the mass 
of the cluster.

For our analysis, we have used the "red arm" of the UVES spectra
which contains the Calcium triplet (CaT) lines around 8500\AA . 
Figure~\ref{fig:cat} shows a representative example spectrum which also
demonstrates a systematic effect which complicates the simple
approach outlined above, in particular if the cluster is dominated
by a young stellar population which contains a significant number of
supergiant stars. In this case, the correct mix 
of template stars is crucial, because the CaT lines in supergiant stars 
are strong enough to be systematically broadened due to their 
damping wings. If not properly reflected in the mix of template
stars used for comparison to the cluster spectrum, this can lead 
to a significant overestimate of the velocity dispersion, and hence 
the cluster mass which scales with $\sigma^2$. 
While the population synthesis methods described in \S~3 can provide 
important constraints on the correct template mix, it is often
more reliable (provided the signal-to-noise of the data is high enough)
to only use (much weaker) metal lines which originate in the photosphere
of cool supergiants (with spectral type G-M). These lines are never 
strong enough for their 
profiles to be affected by damping, and have intrinsic widths that are 
small enough to allow measurement of the typical dispersion values in 
massive stellar clusters (of order 20-30 km/s) (or a full discussion
of this issue, see Ho \& Filippenko 1996a,b).
Figure~\ref{fig:cat} indicates the range of values
for $\sigma$ that are derived by fitting the full spectral range
(incl. the CaT lines, case 1) and the metal lines only (case 2).

While we will discuss the detailed methodology and error analysis
in a forthcoming paper (Walcher \ea\ 2004), our preliminary results 
indicate that the typical masses of {\it present-day} NCs 
are a few $10^6\,\msun$. While this is significantly higher than 
the average globular cluster mass, it is not necessarily inconsistent 
with the notion that at least some GCs might once have been NCs, because 
gas inflow onto the cluster was presumably stopped when their host 
galaxy was disrupted. In any case, it is clear from the fact that the 
sizes of both cluster families are comparable (see \S~2) that NCs must 
have a very high mass density. 
%
%
\section{Is there fuel for repetitive nuclear starbursts?}
A natural question to ask is whether the latest-type spirals 
contain a sufficiently large reservoir of molecular gas in 
their central regions to support repetitive nuclear starbursts.
To this end, we have surveyed 47 late-type spirals (most of them 
taken from the sample of paper~1) with the IRAM 30m-telescope
to quantify the amount of CO in the central $20\arcsec$ which 
corresponds to about 1.5 kpc at the median distance of our sample 
(16.7\,Mpc). The results of this survey (B\"oker et al. 2003) can 
be summarized as follows:

\begin{figure}[!ht]  
\plotone{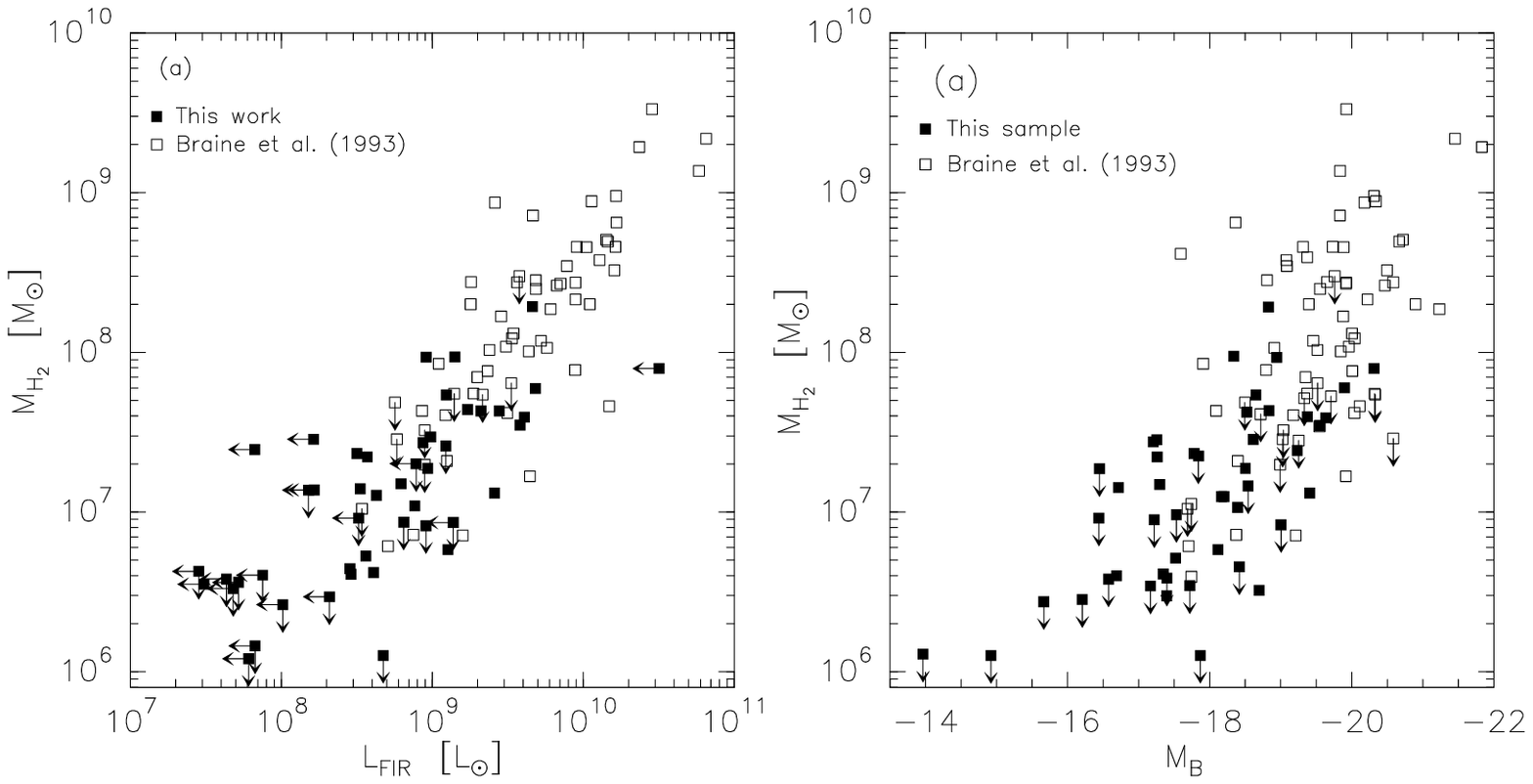}
\caption{H$_2$ mass within the IRAM beam as a function
of the FIR (left) and optical (right) luminosity of the galaxy. 
In both cases, the H$_2$ mass has been calculated from the CO 
luminosity using the standard Galactic conversion factor. While the 
latest-type spirals lie at the faint end of the correlation, they
follow the trend established for more luminous, earlier-type spirals
(B\"oker \ea\ 2003).
}
\label{fig:h2}
\end{figure}
1) For galaxies with CO detections, the median H$_2$ 
mass within the central kpc is $\rm 1.4\times10^7\,M_{\sun}$ 
(assuming the standard Galactic conversion factor), an amount  
that can support a number of modest starburst events 
($\rm \sim 10^5\,M_{\sun}$ in stars). 
At least to first order, this is consistent with the scenario
of a series of ``delta-bursts'' in the nucleus which explains the 
fact that the SED of most clusters is dominated
by a young (less than a few 100 Myrs) population of stars.

2) The latest-type spirals closely follow correlations between
molecular gas content and galaxy luminosity - both at optical and 
far-infrared wavelengths. These correlations, which have been 
established for more luminous, early- and intermediate-type spirals,
are extended to lower luminosities by our sample (see Figure~\ref{fig:h2}). 
This confirms the notion that the latest-type
spirals are ``normal'' spiral galaxies which has already been
inferred from their angular momenta and rotation velocities
(Matthews \& Gallagher 2002).

3) We find a strong separation between our CO detections and 
non-detections in central surface brightness of the stellar disk, 
$\mu_I^0$.
We detect 93\% of galaxies with $\mu_I^0 < 19$ mags arcsec$^{-2}$, 
but only 8\% of galaxies with $\mu_I^0 > 19$ mags arcsec$^{-2}$. 
While our observations are not
sensitive enough to establish a direct correlation between the two
quantities, our results fit in well with the detection rate of 
low-surface brightness (LSB) galaxies (O'Neil et al. 2003) which 
have even lower surface brightness disks than the galaxies
in our sample. This suggests that the stellar mass distribution of the
galaxy disk is an important indicator for the central accumulation
of molecular gas.

4) In contrast to studies of earlier-type spirals (e.g. Sakamoto 1999),
our observations provide little evidence for an enhanced 
molecular gas mass in the centers of barred galaxies relative to 
unbarred ones. It thus remains unclear what the mechanism
for regulating the gas inflow into the central kpc is.
%
%
\section{Summary}
In ellipticals and ``classical'' spirals, phenomena that indicate 
nuclear activity such as AGN and massive black holes 
have received much attention over the past decades. Only 
recently, however, has it become evident that even in bulgeless,
``pure'' disk galaxies which are generally devoid of any obvious 
signs for nuclear activity,
the galaxy center is a ``special'' location in that it is occupied by
a massive, compact, and often young stellar cluster. These nuclear
star clusters (NCs) have sizes and luminosities that are very similar 
to those of young super star clusters (SSCs) found in the disks of 
starburst galaxies. However, NCs most likely have a different, more 
complex, formation history, as can 
be inferred from population synthesis age-dating of their spectra as 
well as estimates of their masses from stellar dynamical modelling.

Unless one is willing to believe that we live in a special time,
the fact that many nuclear star clusters are young suggests
that they experience repetitive ``rejuvenation'', most likely 
due to infall of molecular gas into the central few pc and 
associated star formation. This is not implausible, because
bulgeless spirals are in many ways normal spirals. In particular,
their molecular gas content follows the same scaling relation with
galaxy luminosity as earlier-type spirals, and typical molecular 
gas masses in their central kpc are of the order $10^7\,\msun$.

While the ``duty cycle'' for nuclear starbursts still needs to
be reliably established - which is one of the goals of our project - 
the possibility of repetitive nuclear cluster formation in late-type
spirals has interesting consequences for galaxy evolution scenarios, 
both with respect to morphological classification (i.e. Hubble type) and 
nuclear activity (i.e. black hole growth). 


\end{document}